
\input harvmac
\input epsf.tex
\noblackbox
\overfullrule=0pt
\def\Title#1#2{\rightline{#1}\ifx\answ\bigans\nopagenumbers\pageno0\vskip1in
\else\pageno1\vskip.8in\fi \centerline{\titlefont #2}\vskip .5in}

%
%
%

\def\[{\left [}
\def\]{\right ]}
\def\({\left (}
\def\){\right )}

\def\p{\partial}

%

%

\lref\ascv{A. Strominger and C. Vafa, ``{\it On the Microscopic
Origin of the Bekenstein-Hawking Entropy}'', Phys. Lett. B379 (1996)
99,
 hep-th/9601029.}

\lref\taylor{W. Taylor, ``{\it Adhering 0-Branes to 6-Branes and 8-Branes}'',
hep-th/9705116.}

\lref\hull{C. Hull and P. Townsend,
Nucl. Phys. B438 (1995) 109,
hep-th/9410167.}

\lref\cremer{E. Cremmer and B. Julia, Nucl. Phys.  B159 (1979)
141.}


\lref\cveticfifth{ M. Cvetic and D. Youm,  hep-th/9512127;
M. Cvetic and A. Tseytlin, Phys. Rev. D53 (1996) 5619, hep-th/9512031.
}

\lref\cvetichull{ M. Cvetic and C. Hull, Nucl. Phys. B480 (1996) 296,
hep-th/9606193.}

\lref\cveticgaida{ M. Cvetic and I. Gaida, hep-th/9703134.}

\lref\bll{V. Balasubramanian, F. Larsen and R. Leigh,
hep-th/9704143.}

\lref\sfst{
L. Andrianopoli, R. D'Auria, S. Ferrara,
P. Fr{\'e},
    M. Trigiante, ``{\it  R-R Scalars,
U-Duality and Solvable Lie Algebras}'',
 hep-th/9611014.}

\lref\sftwo{L. Andrianopoli, R. D'Auria, S. Ferrara,
``{\it  U-Invariants, Black-Hole Entropy and Fixed Scalars}'',
 hep-th/9703156.
}

\lref\kol{R. Kallosh and B. Kol,
Phys. Rev. D53 (1996) 5344,  hep-th/9602014.
}

\lref\sfrk{S. Ferrara and R. Kallosh,
Phys. Rev. D54 (1996) 1525,
hep-th/9603090.
}

\lref\susyandcosmic{R. Kallosh, A. Linde, T. Ortin, A. Peet, A. Van
Proeyen,  Phys. Rev. D46 (992) 5278, 
hep-th/9205027.}

\lref\jp{J. Polchinski, private communication.}



\lref\hms{G. Horowitz, J. Maldacena and A. Strominger,
Phys. Lett. B383 (1996) 151,
hep-th/9603109.}

\lref\ascv{A. Strominger and C. Vafa, Phys. Lett. B379 (1996) 99,
hep-th/9601029.
}

\lref\cama{ C. Callan and J. Maldacena, Nucl. Phys. B475 (1996) 645,
hep-th/9602043.}


\lref\cremerother{ E. Cremmer in ``{\it Supergravity}'', ed. by
S. Ferrara and J. Taylor, p. 313;
B. Julia in ``{\it Superspace \& Supergravity}'', ed. by S. Hawking
and
M. Rocek, Cambridge (1981) p. 331. }

\lref\sentorus{A. Sen,  Nucl. Phys. B440 (1995) 421, hep-th/9411187.}

\lref\harrison{ H. Sheinblatt, ``{\it Statistical Entropy of an Extremal
Black Hole with  0- and 6-Brane Charges}'', hep-th/9705054.}

\lref\polchinskinotes{J. Polchinski,``{\it TASI lectures on D-branes}'',
hep-th/9611050.}

\lref\senmonopole{A. Sen, ``{\it Kaluza-Klein Dyons in String Theory}'',
 hep-th/9705212.}

\lref\berkooz{M. Rozali, ``{\it Matrix theory and U-duality in seven
dimensions}'', hep-th/9702136; M. Berkooz, M. Rozali and N. Seiberg,
``{\it Matrix Theory Description of M-theory on $T^4$ and $T^5$}'', 
hep-th/9704089.}

\lref\bfss{ T. Banks, W. Fishler,  S. Shenker and  L.  Susskind,
``{\it M theory as a Matrix Model: A Conjecture}'', hep-th/9610048.}

\lref\lennyori{ L. Susskind, ``{\it T duality in M(atrix) theory and S 
duality
in field theory}'', hep-th/9611202;    O. Ganor, S. Ramgoolam and W.
 Taylor,
``{\it Branes, Fluxes and Duality in M(atrix) Theory}'', hep-th/9611202.}

\lref\slansky{ R. Slansky, Phys. Rep.  79 (1981) 1 (see pages 108, 112, 113).}

\lref\fks{S. Ferrara, R. Kallosh 
and A. Strominger, Phys. Rev. D52 (1995) 5412, hep-th/9508072.}

\lref\sfrktwo{ S. Ferrara and R. Kallosh, Phys. Rev. D54 (1996) 1514,
hep-th/9602136.}

\lref\normal{S. Ferrara, C. Savoy, B. Zumino, Phys. Lett. 100B (1981)
393.}

\lref\duff{M. Duff, J. T. Liu and J. Rahmfeld, Nucl. Phys. B459 (1996)
125, hep-th/9508094.}

\lref\ortin{ R. Kuhri  and  T. Ortin, Phys. Lett. B373 (1996) 56, 
hep-th/9512178. }

%
\Title{\vbox{\baselineskip12pt
\hbox{hep-th/9706097}\hbox{ RU-97-35}\hbox{ CERN TH 97/112}
}}
{\vbox{
{\centerline { Branes, central charges and U-duality invariant }}
{\centerline {
BPS conditions
 }}
  }}
\centerline{Sergio Ferrara\foot{FERRARAS@vxcern.cern.ch} and 
 Juan Maldacena\foot{malda@physics.rutgers.edu
}}
\vskip.1in
\centerline{\it$^1$ CERN Theoretical Division, CH 1211 Geneva 23, Switzerland
}
\vskip.1in
\centerline{\it$^2$ Department of Physics and Astronomy, Rutgers University,
Piscataway, NJ 08855, USA}
\vskip.1in
\vskip .5in

\centerline{\bf Abstract}

In extended supergravity theories there are $p$-brane solutions preserving 
different numbers of supersymmetries, depending on the charges, the
spacetime dimension and the number of original supersymmetries (8, 16
or 32).
 We find  U-duality invariant
conditions on the quantized charges which specify the number of supersymmetries
preserved with a particular charge configuration.
These conditions relate U-duality invariants to the picture of
intersecting
branes.
The analysis is carried out for all extended supergravities with
16 or 32 supersymmetries in  various dimensions.

 \Date{}

%

\newsec{ Introduction }

Extended supergravity theories contain BPS black hole solutions which
preserve some supersymmetries. 
Given a generic charge configuration we can find an extremal black
hole solution, extremal in the sense of the cosmic censorship bound,
i.e. the black hole solution with a mass that saturates the 
bound coming from demanding that there be  no naked singularities. 
 In some cases the extremal black hole is also BPS \susyandcosmic\ and in some
others it is not BPS. Even in the case when the extremal
solution is supersymmetric it can preserve different
numbers of supersymmetries. 
The charges  transform under a  group $G$
characteristic of the supergravity theory. We show how
different cases are separated by $G$-invariant conditions
on the charges.

Maximally extended supergravities in $d$ spacetime dimensions are
the low energy limit of type II (A or B) string theory
 compactified on a 
10-$d$ dimensional torus.
The classical supergravity theories are formulated in terms
of  an underlying 
non-compact group $G$
\cremer  \cremerother\ which is $G=E_{11-d}$, i.e.
$E_7$ in four dimensions, $E_6$ in five dimensions, 
$E_{5} = S0(5,5) $ in six dimensions, $E_{4} = SL(5,R)$ in seven
dimensions, $E_3 = SL(3,R)\times SL(2,R)$ in eight dimensions
and $E_2 = SL(2,R) \times O(1,1) $ in nine dimensions. 
These  theories contain some  Abelian $p+1$-form potentials which,
for each $p$, 
 arrange themselves into multiplets of the  group $G$.
They also have a large number of scalar fields which live in the
coset space  $G/H$, where $H$ is
a maximal compact subgroup of the non-compact group $G$.
Even though the  symmetry of the theory is
only $H$ the charges transform under the group $G$. 
In the full quantum theory, charges are quantized and the duality 
symmetry is broken to  a discrete subgroup $G(Z)$, which is 
the U-duality group \hull . 
Classical supergravity solutions correspond to the limit of large
values of integer quantized charges. In this 
case the action of $G(Z)$ becomes almost continuous, so we will
consider the action of the continuous group. 
We normalize the charges so that they become  integers 
in the quantum theory. 
The  various $p+1$-forms
couple to $p$-dimensional objects
and we always consider $p< d-3 $ so that
fields decay fast enough at infinity, enabling the action of the 
various (super)symmetry generators on the configuration to be  well
defined. If this condition is not satisfied the geometry 
 will not be asymptotically
Minkowski in the presence of a brane.
 Configurations with $p$ and $d-p - 4$
 are related by Dirac 
electric-magnetic duality. The corresponding charges transform in
the gradient and contragradient representations of the
group $G$, except for the  
 case  $p = {d-4 \over 2}$ where the same representation includes both
electric and magnetic  objects. 
The commutator of two supersymmetries  contains several  ``central charge
matrices''. These central charge matrices   depend on the charges and
on the moduli of the theory, which are the values of the scalar fields
at infinity. 
By using a transformation in the group $H$ it is possible to 
reduce the central charges  into a normal form where they are
``diagonal'' \normal . BPS states with enhanced supersymmetry restrict
the eigenvalues of the central charge matrices, giving constraints
on the charges. 
 The central charges in the normal frame
still depend on the moduli. These are residual $G/H$ transformations 
that keep the diagonal structure of the matrix in the normal
frame. These tranformations are just some number of $O(1,1)$
rotations.
We can view this choice of the normal frame as choosing
a particular background of a square torus where the charges
are aligned in a simple way on the torus. 
The residual transformations are related to the possibility of
changing
the size of the torus, etc. \sentorus .

In the case of extremal BPS black hole solutions it is possible
to write
the entropy in terms of the charges in a 
U-dual fashion. This makes use of particular quartic \kol\ and cubic
\jp \sfrk\
invariants in 4 and 5 dimensions respectively.
This is the situation when the charges are generic.
There are however some charge configurations for which the area
of the black hole horizon is zero, and also some configurations
which preserve a larger number of supersymmetries.
There are also configurations that  do not preserve any
supersymmetries. 
In different dimensions the number of possible preserved supersymmetries
can be calculated as follows. A localized particle-like 
configuration breaks the Lorentz group into the little group 
$SO(d-1)$. The preserved supersymmetry has to be 
in a representation of $SO(d-1)$. For $d$ = 4,5 the spinor representation
has 4 real components and for $d$ = 6,7,8,9 the spinor representation has
8 real components.  So depending on the number of original
supersymmetries (which is 32 in the maximal case) we have
different numbers of possible preserved supersymmetries: 1/2, 1/4, 1/8
depending on the dimension. 
For example, in $d \ge 6$ we can have only 1/2 or 1/4 BPS solutions.

We write down U-duality invariant expressions which separate the
various
cases. We also argue that one can choose a ``basis'' for the
charges in which each element by itself breaks 1/2 of the
supersymmetry
and that, taken together, they break more supersymmetries.
This ``basis'' has a representation in terms of intersecting
branes.

We can 
 decompactify the $d$-dimensional theory into
a $d-1$ dimensional one by letting one of the radii of the torus
go to infinity.
The duality group decomposes as 
$E_{11-d} \to E_{10-d} \times O(1,1)$ where the $O(1,1)$ is
related to the  $T$-duality symmetry that is lost when a circle
becomes infinite\foot{In 
the quantum theory $O(1,1) \to Z_2$.}.
 It will be useful to analyze the behaviour
of the representations under this decomposition.

It is also instructive  to decompose $E_{11-d}$ under $S,T$ duality.
The decomposition reads
$E_{11-d} \to  O(1,1) \times O(10-d,10-d)$ for $d \ge 5$ and 
 $E_7 \to  SL(2,R) \times O(6,6)$ for $d=4$ \sfst .
This decomposition separates NS and R charges in string theory.

This paper is organized as follows:
In section 2 we analyze maximal supergravities (32 supersymmetries)
in various dimensions and for different extended objects. The 
conditions for a state to preserve different numbers of
supersymmetries are presented. 
In section 3 the analysis is extended to theories with
16 supersymmetries in $d=4,5$, such as heterotic on
$T^6, ~T^5$ or the dual type II on $K3\times T^2$, etc.

\newsec{Maximal supergravity in various dimensions, $4\le d \le 9$ }

\subsec{ Five dimensions}

We start with the five-dimensional case. We have 27 Abelian gauge
fields which transform in the fundamental representation of
$E_6$. The electrically charged objects are point-like and the 
magnetic duals are one-dimensional, or string-like. 
The first invariant of $E_6$ is the cubic invariant 
$I_3 = T_{ijk} q^i q^j q^k $.
In fact, the entropy of a  black hole with charges $q^i$ is
proportional to $\sqrt{I_3}$ \jp \sfrk .
We will see that a 
 configuration with $I_3 \not = 0$ preserves 1/8 of the supersymmetries.
If $I_3 =0 $ and ${ \p I_3 \over \p q^i } \not = 0$  then it preserves
1/4 of the supersymmetries, and finally if ${ \p I_3 \over \p q^i }
=0$
(and the charge vector $q^i$ is non-zero), the configuration preserves
1/2 of the supersymmetries.
We will show this by choosing a particular basis for the charges, the
 general result following by U-duality.

In five dimensions the compact group $H$  is $USp(8).$\foot{
We choose our conventions so that $USp(2) = SU(2)$.}
In the commutator of the supersymmetry generators we have
a central charge matrix $Z_{ab}$ which can be brought to a normal
form by a $USp(8)$ transformation. In the normal form the central
charge matrix  can be written
as
\eqn\normal{
e_{ab} = \pmatrix{ s_1 + s_2 - s_3  & 0& 0& 0\cr 0& 
s_1+s_3 -s_2 &0 &0 \cr
0& 0 & s_2 + s_3 - s_1  & 0 \cr 0 & 0 & 0 & - (s_1 + s_2 + s_3)  } 
\otimes \pmatrix{ 0 & 1  \cr  -1 & 0}
}
we can order $s_i$ so that $s_1 \ge s_2 \ge |s_3|$.
The cubic invariant, in this basis, becomes \sfrk 
\eqn\cubic{
I_3 = T_{ijk} q^i q^j q^k = s_1 s_2 s_3 ~.
}
Even though the eigenvalues $s_i$ might depend on the moduli, the
invariant \cubic\ only depends on the quantized values of the
charges.
We can write a generic charge configuration as
$U e U^t $,  where $e$ is the normal frame as above, and the 
invariant will then be \cubic . 
There are three distinct possibilities 
\eqn\possib{\eqalign{
I_3 \not = 0 & ~~~~~~~~ s_1,~s_2,~ s_3 \not = 0 \cr
I_3 =0 ,~~~{ \partial I_3 \over \partial q^i } 
\not =0 &~~~~~~~~ s_1,~s_2 \not =0,
~~~~~~~ s_3 =0
\cr
I_3 =0 ,~~~~{ \partial I_3 \over  \partial q^i } =0& ~~~~~~~~~ s_1\not =0,
~~~~~~~s_2, ~ s_3 =0
}}
Taking the case of type II on $T^5$ we can 
 choose the rotation in such a way that, for example,   $s_1 $
corresponds 
to solitonic five-brane charge, $s_2$ to fundamental string winding
charge along some
direction and $s_3$ to Kaluza-Klein momentum along the same direction. 
We can see that in this specific example the three possibilities in
\possib\
break 1/8, 1/4 and 1/2 supersymmetries.
This also shows that one can generically choose a basis for the
charges
so that all others are related by U-duality. 
The basis chosen here is the S-dual of the $D$-brane basis usually chosen for
describing black holes in type II B on $T^5$ \ascv \cama . 
All others are related by U-duality to this particular choice.
The sign of the invariant \cubic\ is not important  since it changes
under a CPT transformation.

In five dimensions there are also string-like configurations
which are 
the magnetic duals of the  configurations considered here. They 
transform in the contragradient 27 representation and the solutions preserving 
1/2, 1/4, 1/8 supersymmetries are characterized in an analogous way. 
We could also have configurations where we  have both point-like  and
string-like charges. If
the point-like charge is uniformly distributed along the string,
it
is more natural to consider this configuration as a point-like
object 
in $d=4$ by dimensional reduction.

It is useful to decompose the U-duality group into the T-duality 
group and the S-duality group \sfst . The decomposition reads $E_6 \to
O(5,5) \times O(1,1)$,  leading to 
\eqn\decompts{
{\bf 27 } \to {\bf 16}_{1} + {\bf 10 }_{-2}+ {\bf 1}_{4} ~.
}
The last term in \decompts\ corresponds to the NS five-brane charge.
The ${\bf 16}$ correspond to the D-brane charges and the 
${\bf 10 }$ correspond to the 5 directions of KK momentum and
the 5 directions of fundamental string winding, which 
are the charges that explicitly appear in string perturbation 
theory.
The cubic invariant has the decomposition
\eqn\cubicdects{
({\bf 27})^3 \to {\bf 10}_{-2}\ {\bf 10}_{-2}\ {\bf 1}_{4} +
{\bf 16}_{1}\ {\bf 16}_1\ {\bf 10}_{-2} ~.
}
This is  saying that in order to have
a non-zero area black hole we must have three NS charges
(more precisely some ``perturbative'' charges and a solitonic
five-brane); or  we can have two D-brane charges and one NS charge.
In particular, it is not possible to have a black hole
with a non-zero horizon area with purely D-brane charges.

Notice that the non-compact nature of the groups is crucial in
this classification.

\subsec{Four dimensions}

In four dimensions the duality group is $E_7$ and the charges
transform
in the 56 representation of $E_7$. In this case electric and magnetic
charges are all point-like and are included in the same representation
of the duality group.
The invariant is quartic $I_4 = T_{ijkl} q^i q^j q^k q^l $ \kol
\cvetichull \cveticgaida\
and it can also be expressed in terms of the central charge 
matrix $Z_{AB}(q,\phi)$. Of course, the dependence
on the scalar fields drops out from the  expression for $I_4$.
Again, by performing an $SU(8)$ transformation
 one can choose the charges in the normal frame form
\eqn\normalf{
Z_{ab} =  \pmatrix{ z_1 & 0& 0& 0\cr 0& z_2 &0 &0 \cr
0& 0 & z_3 & 0 \cr 0 & 0 & 0 & z_4 } \otimes \pmatrix{
0& 1 \cr -1 & 0 } ~, 
}
where $z_i=\rho_i e^{i\varphi_i}$ are complex. Actually the relative
phases of $z_i$ can be changed,
 but the overall phase $\varphi = \sum \varphi_i$
cannot be removed by an $SU(8)$ transformation; it 
 is related to an extra parameter in 
the class of black hole solutions \cveticfifth .
In this basis the quartic invariant takes the form \kol
\eqn\qarticphase{\eqalign{
I_4 &= \sum_i |z_i|^4 -2 \sum_{i<j} |z_i|^2 |z_j|^2 + 4(z_1z_2z_3z_4 +
\bar z_1 \bar z_2 \bar z_3 \bar z_4 ) \cr 
& = (\rho_1 + \rho_2 + \rho_3 + \rho_4)
(\rho_1 + \rho_2 - \rho_3 - \rho_4  )
(\rho_1 - \rho_2 + \rho_3 -  \rho_4  )
(\rho_1  - \rho_2 -   \rho_3  + \rho_4  )+ \cr
&+8 \rho_1 \rho_2 \rho_3 \rho_4 ( \cos  \varphi  -1 ) ~.
}}
It was shown in \sfrk \sftwo\ that for a 1/8 supersymmetric solution
\eqn\bhcondtion{
M_{BPS}^4(\phi_h, q) = I_4 ~,~~~~~~~~~z_1(\phi_h,q)\not = 0,~~~~~~~~~
z_i(\phi_h, q) =0,~~i=2,3,4 ~,
}
where $\phi_h$ are the values of the moduli at the horizon, given by ratios of
the 
quantized charges \fks \sfrktwo . 
This implies that $I_4 \ge 0$ for a BPS solution and if $I_4 <0$ then
the extremal solution is not  BPS (as opposed to $d=5$ where the
sign of $I_3$ was not important). We also see that at the horizon 
 we can choose $\varphi_h =0$.

The condition for a 1/4 BPS state is \sfrk
\eqn\onequarter{
|z_1(\phi, q) | = | z_2(\phi, q)|~,~~~~~~~~~~~
|z_3(\phi,q)| =|z_4(\phi,q)| .
}
This happens when ${ \partial I_4 \over \partial z_i} =0$, 
in the normal frame this implies, in particular, $\phi =0$ and 
also $ \rho_1 = \rho_2 $, $\rho_3 =\rho_4$.
We then  see  that there is no extra phase if the configuration
preserves at least 1/4 supersymmetry.

If the state preserves 1/2 of the supersymmetries the
condition on the central charges is
$\phi =0, ~~\rho_1 =\rho_2 =\rho_3 =\rho_4 $ \sfrk .
This translates into the condition that
the second derivatives of $I_4$ projected on the 
adjoint representation vanish, 
\eqn\projectadj{
\left. {\partial^2 I_4 \over \partial q^i \partial q^j }\right|_{Adj.} 
\sim T_{ijkl} q^k q^l|_{Adj} =0~.
} 
There is no constraint on the ${\bf 1463}$ representation of
$E_7$ present in the above symmetric polynomial \slansky . 
Note that under $SU(8)$: ${\bf 1463 } = {\bf 1} + \cdots $, where
the singlet is the 1/2 BPS mass which can be extracted from $I_4$ as
follows 
\eqn\masa{
M^2_{BPS} = - {1\over 8} \sum_i {\partial^2 I_4 \over
\partial z_i \partial \bar z_i } = \rho^2
}

If the phase vanishes, $I_4$ becomes
\eqn\quartic{
I_4 =  T_{ijkl} q^i q^j q^k q^l = s_1 s_2 s_3 s_4~,
}
where we have defined $s_i$ by 
\eqn\eass{\eqalign{
s_1 = & \rho_1 + \rho_2 + \rho_3 + \rho_4 \cr
s_2 = & \rho_1 + \rho_2 - \rho_3 - \rho_4 \cr
s_3 = & \rho_1 - \rho_2 + \rho_3 - \rho_4 \cr
s_4 = & \rho_1 - \rho_2 - \rho_3 + \rho_4 ~.
}}
and we order 
 the $s_i$ so that $s_1 \ge s_2 \ge s_3 \ge |s_4| $. 
Now the distinct possibilities are
\eqn\possibfour{\eqalign{
I_4 \not =0 & ~~~~~~~~~~~~~s_1,s_2,s_3,s_4 \not = 0 \cr
I_4 =0, ~~~~~ {\p I_4 \over \p q^i } \not = 0 &~~~~~~~~~~~~~
 s_1,s_2,s_3  \not = 0,~~~~~ s_4 =0 \cr
{\p I_4 \over \p q^i }  = 0 , ~~~~~ 
\left. {\p^2 I_4 \over \p q^i \p q^j }\right|_{Adj E_7} \not=0
& ~~~~~~~~~~~~~ s_1,s_2 \not = 0,~~~~~ s_3, s_4 =0 \cr
\left. {\p^2 I_4 \over \p q^i \p q^j }\right|_{Adj E_7}  =0 & ~~~~~~~~~~~~~ 
s_1 \not = 0,~~~~~~ s_2 , s_3, s_4
=0
}}

We  can choose a basis of four charges $(q_i)_I$, $I=1,2,3,4$ such
that for each element  only $s_I \not = 0$ and $s_i =0, ~ i\not =I$.
Any charge vector is then related by an $E_7$ rotation to a 
configuration with only these four charges if the phase vanishes. 
An example of this  would be a set of 
four D-three-branes oriented along $456,~678,~894,~579$ (where the order
of the three numbers indicates the orientation of the brane).
Note that in choosing the basis the sign of the D-3-brane charges is
important; here they are chosen such  that taken together with positive
coefficients  they
form a BPS object.
The first two possibilities in \possibfour\ preserve
1/8 of the supersymmetries, the second 1/4 and the last 1/2.
It is interesting that there are two types of 1/8 BPS solutions.
In the supergravity description, the difference between them is
that the first in \possibfour\
has non-zero horizon area.
If $I_4 < 0$ the solution is
not BPS.
 This case corresponds, for
example, to changing the sign of one of the three-brane charges 
discussed above. By U-duality transformations we can relate this
to configurations of branes at angles such as in \bll .

Going from four to five dimensions
it is natural to decompose the $E_7 \to E_6 \times O(1,1)$ where
$E_6 $ is the duality group in five dimensions
and $O(1,1)$ is the extra $T$ duality that appears when 
we compactify from five to four dimensions.
According to this decomposition the representation
breaks as $ {\bf 56 } \to {\bf 27 }_1 + {\bf 1}_{-3} + 
{\bf 27' }_{-1} + {\bf 1}_3 
$
and the quartic invariant becomes 
\eqn\quartdecomp{
{\bf 56 }^4 = ({\bf 27 }_1 )^3 {\bf 1}_{-3} +
({\bf 27' }_{-1} )^3 {\bf 1}_{3} +  
{\bf 1}_{3} {\bf 1}_{3} {\bf 1}_{-3} {\bf 1}_{-3} +
{\bf 27 }_{1}{\bf 27 }_{1}{\bf 27' }_{-1}{\bf 27' }_{-1} +
{\bf 27 }_{1}{\bf 27' }_{-1}{\bf 1}_{3} {\bf 1}_{-3}
}
The ${\bf 27}$  comes from point-like charges in five dimensions
an the ${\bf 27'}$  comes from string-like charges.

Decomposing  the U-duality group into T- and
S-duality
groups,  $E_7 \to SL(2,R) \times O(6,6)$ we find 
${\bf 56 } \to ({\bf 2}, {\bf 12} ) + ({\bf 1 }, {\bf 32 })$
where the first term corresponds to NS charges and the second term
to D-brane charges.
Under this decomposition the quartic invariant \quartic\
becomes
${\bf 56}^4 \to {\bf 32 }^4 + ({\bf 12 .  12'})^2 + { \bf 32}^2 .
{\bf 12 . 12'} $. This means that we can have configurations
with a non-zero area that carry only D-brane charges, or only 
NS charges or both  D-brane and  NS charges. We can then express the
charges as $e_{AB} = (v_i^\alpha , S^a)$, where $\alpha =1,2$,
$i=1,...,12$ is the vector index and $a =1, 32$ is the spinor index.
In order to gain some light on the conditions in \possibfour\
involving
the projections on the adjoint we decompose them according to
the S-T-duality groups. The adjoint representation of $E_7$ 
decomposes as 
${\bf 133} \to ({\bf 3}, {\bf 1 })+({\bf  1 }, {\bf 66 })+({\bf 2},
{\bf 32 })$.
The last condition in \possibfour\ becomes, with this decomposition,
\eqn\projection{\eqalign{
{ \p^2 I_4 \over \p S^a \p S^b } (\gamma^{ij})_{ab} + &
{\p^2 I_4 \over \p e^{\alpha}_i \p e^{\beta}_j } \epsilon_{\alpha
\beta}
=0 ~,\cr
{ \p^2 I_4 \over \p S^a \p e^\alpha_i } (\gamma_i )_{ab} =0~,
&~~~~~~~~~~{\p^2 I_4 \over \p e^{\alpha}_i \p e^{\beta}_j } \eta_{ij}
=0~.
}}
where $\gamma^i$ are the $O(6,6)$ gamma-matrices. 
   
 An interesting case where $I_4$ is negative
 corresponds to a configuration carrying electric and magnetic 
 charges under the same gauge group, for example a 0-brane plus
 6-brane configuration which is U-dual to a KK-monopole and 
plus KK-momentum \ortin \harrison . This case corresponds
to $z_i = \rho e^{i\varphi/4}$ and the phase is 
$\tan \varphi/4 = e/g $ where $e$ is the electric charge
and $g$ is the magnetic charge. Using \qarticphase\  we 
find that $I_4<0$ unless the solution is purely electric or
purely magnetic.  In \polchinskinotes\ it was suggested that
$0+6$ does not form a supersymmetric state. 
Actually it was shown in 
 \taylor\ that a 0+6 configuration can be T-dualized into
 a non-BPS configuration of four intersecting D-3-branes.
Of course, $I_4$ is negative for both configurations.
Notice that even though these two charges are Dirac dual (and U-dual) they
are not S-dual in the sense of filling out an $SL(2,Z)$ multiplet. 
In fact, the KK-monopole forms an $SL(2,Z)$ multiplet with a fundamental
string winding charge under S-duality \senmonopole .

\subsec{Six dimensions}

In this case the duality group is O(5,5) and  
 we have vector fields and two form field potentials.
The vector fields couple to point-like configurations and 
their magnetic duals to two-dimensional configurations.
The two-form potentials and their magnetic duals both couple to 
one dimensional string-like objects.
All point-like charges belong to the  ${\bf 16}$ representation
(spinor of O(5,5))
while one-brane charges belong to the ${\bf 10}$ (vector of O(5,5)).

Going from five to six by decompactifying one dimension
leads to the decomposition $ E_6 \to O(5,5)\times O(1,1)$ and
the representations decompose as 
\eqn\decomp{\eqalign{
{\bf 27 } \to & {\bf 16}_{1} + {\bf 10 }_{-2}+ {\bf 1}_{4} \cr
{\bf 27' } \to & {\bf 16'}_{-1} + {\bf 10 }_{2}+ {\bf 1}_{-4} ~. 
}}
The ${\bf 1}_4 $  corresponds to  $KK$ momentum
and the ${\bf 1}_{-4}$ to KK monopole charge. From
 group theory, this is the same decomposition as in \decompts\
but the interpretation is different.
The cubic invariant has the decomposition
\eqn\cubicdec{
({\bf 27})^3 \to {\bf 10}_{-2}\ {\bf 10}_{-2}\ {\bf 1}_{4} +
{\bf 16}_{1}\ {\bf 16}_1\ {\bf 10}_{-2}~.
}

Solutions carrying one-brane charge can preserve $1/2$ or $1/4$ 
of the supersymmetries according to whether
the vector ${\bf 10}$ is null or not, respectively. 
Similarly a point-like solution, characterized by the spinor $S^a$,
can preserve 1/2 or 1/4 according to whether
$S^a \gamma^\mu_{ab} S^b $ is zero (as a vector)  or not,
 respectively.
We see that both conditions are U-duality invariant.

A one-dimensional solution can also carry ``zero-dimensional'' charge;
this charge can be spread uniformly along the string. 
These configurations can break more supersymmetries, leaving only
1/8,
when the invariant ${\bf 16}_{1}\ {\bf 16}_1\ {\bf 10}_{-2} $ is non-zero,
and they have a natural interpretation as black holes in $d=5$.

\subsec{Seven dimensions}

The duality group is $SL(5)$.
We have again  vector potentials and  two-form potentials.
So we have point-like configurations whose
magnetic duals are three-branes and  stringlike configurations
whose magnetic duals are two-branes.
In going from six to seven dimensions  the duality group breaks as
$O(5,5) \to SL(5) \times O(1,1)$ and the representations
${\bf 10 } \to {\bf 5}_{2}  + {\bf \tilde 5 ' }_{-2}$,  
$ {\bf 16} \to {\bf 10 }_{-1} + {\bf 5'}_{3} + {\bf 1 }_{-5}
$.
The point-like charges belong to the antisymmetric tensor ${\bf 10}$ 
and the 
string-like solutions to the ${\bf 5'}$ (vector), and the three-branes
and two-branes to ${\bf 10'}$ and ${\bf 5}$, respectively.
Going to six dimensions the ${\bf \tilde 5'}$ and ${\bf  5'}$
correspond to leaving the string unwrapped or to wrapping it
in the extra circle,  respectively.
In the type IIA the point-like charges would
be the 3 directions of KK momentum, D0-brane charge, 3 directions
of fundamental  string winding, and 3 possible D2-brane wrapping modes.
The string like-charges are 1 D4-brane, 3 D2-branes and one
fundamental
string. The two- and three-brane charges are the magnetic duals of 
these.

The invariants break as
\eqn\invbrea{\eqalign{
{\bf 10}^2 \to & {\bf 5}_2 \ {\bf \tilde 5'}_{-2} \cr 
{\bf 16}\ {\bf 16}\ {\bf 10} \to & {\bf 10}_{-1}\ {\bf 5'}_3\ {\bf
\tilde 5'}_{-2} +
{\bf 5}_2\ {\bf 5'}_3\ {\bf 1}_{-5} + {\bf 10}_{-1}\ {\bf 10}_{-1}\ 
{\bf 5}_{2}~.
}}

We see that there is no quadratic condition we can impose on a
${\bf 5 }$; this is related to the fact that all one-dimensional 
 configurations
with only string charges break 1/2 of the supersymmetries
 (a fundamental string ending on
a D-brane would preserve 1/4 
 but the configuration  would have to extend along two different directions
in space). On the other hand
we can have point-like solutions preserving 1/2 or 1/4 of the
supersymmetries
according to whether the
 $ \epsilon^{ijklm}T_{ij}T_{kl}$ is zero or not
respectively ($T_{ij}$ is the ${\bf 10}$ representation).

The M(atrix) theory description involves the (0,2) non-trivial fixed
 point
theory 
describing the world-volume degrees of freedom  of coincident 5-branes
in M-theory. M-theory on $T^4$ is defined by compactifying
this (0,2) theory on $T^5$ \berkooz .
The SL(5,Z) duality symmetry is just
the modular group of a five-torus \berkooz . The (0,2)  theory contains
a  two-form potential with a self dual three-form field strength.
 The poin-tlike charges 
correspond to fluxes along three of the spatial dimensions
of the fivetorus, they are naturally in the {\bf 10} of 
{SL(5)}. The string-like solutions correspond to momentum modes along the
torus.
The five possible directions give the five possible string-like
charges. They represent strings along the longitudinal direction
of the  M(atrix) description \bfss .

\subsec{Eight dimensions}

In eight dimensions the duality group is 
SL(3,Z)$\times$SL(2,Z).
We have point-like configurations and their magnetic 4-brane
duals, string-like configurations and their magnetic 3-brane
duals, and finally two-brane configurations.
The point-like configurations $w_{\alpha b}$ are in 
${\bf 3} \times {\bf 2 } $, the string-like configurations in
${\bf 3'} \times {\bf 1} $ and the dyonic  two-brane in 
${\bf 1}\times {\bf 2}$.

In M(atrix) theory, the description is based on a 3+1 YM theory on
a torus. SL(3,Z) comes from the symmetries of the torus while
SL(2,Z) comes from the S-duality of YM \lennyori . 
The six point-like charges correspond to fluxes in the
YM theory, the string-like charges correspond to momentum modes along the
three-torus. 

It is clear that there are no invariant conditions that select 
1/4 or 1/2 in the case of solutions with purely string charges
or purely two-brane charges. This fits in with the fact that those
configurations can only be 1/2 BPS.

The point-like solutions could be 1/2 or 1/4 BPS according to
whether $\epsilon^{ab} w_{\alpha a} w_{\beta b} $ is zero
or not,
where $w_{\alpha a}$ are the point-like charges. 

\subsec{Nine dimensions}

In nine dimensions the duality group is $SL(2,Z)\times Z_2$. 
The point-like charges belong to ${\bf 2}$ ($v_\alpha$) and ${\bf 1}$ ($v$). 
In the IIB case
they would correspond to the two wrapped strings, the fundamental string
and the D-string, and the KK momentum mode.
The string-like charges are in ${\bf2}$ (the fundamental string and the
D-string), the two-brane charge
is  a singlet ${\bf 1}$ (from the ten-dimensional D3-brane)
 and the rest are the magnetic duals of
the above.

With point-like charges we can preserve 1/2 of the supersymmetries
if $v=0$ or $v_\alpha =0$ (i,e $ v v_\alpha =0$)
and  1/4 if both are non-zero.

\newsec{ Supergravities with 16 supersymmetries}

Now we turn to the discussion of supergravity theories with
16 supersymmetries like $N=4$ in $d=4$. 
We will analyze the $d=4,5$ cases. 
If we take a supergravity theory with $n$ matter multiplets
the duality groups are $SL(2,R)\times O(6,n)$ and
$O(1,1)\times O(5,n)$ respectively. 
If we think of heterotic strings on $T^6$ then $n=22$ and
$SL(2,Z)$ in the S-duality symmetry of $N=4$ four dimensional
heterotic strings.

\subsec{Five dimensions}

The charges form a  vector $Q_i$ under $O(5,n)$ and 
a singlet $Q_H$. There are two invariants
$Q_H$ and $Q^2 $. 
In order for a state to be BPS we need $Q^2 \ge 0$.
If either $Q_H =0$ or $ Q_i =0$ (as a vector) 
1/2 of the supersymmetries are preserved and only 1/4 are 
preserved if both are non-zero.
In addition, when $Q_H Q^2$ is non-zero, the corresponding
configuration gives rise to a black hole with non-zero entropy.
Strominger and Vafa computed   the microscopic entropy 
of these black holes
 using D-branes for a general configuration \ascv .
They did the computation for the type II theory on $K3\times S_1$. 
The charge $Q_H$ corresponds to KK momentum along $S_1$ and
the charges $Q_i$ correspond to $D1$-, $D3$- and $D5$-branes
wrapping along $S_1$ and a 0-cycle, a 2-cycle, and a 4-cycle
on $K3$ respectively. 

\subsec{Four dimensions}

In four dimensions we have electric $Q_i$ and magnetic $P_i$
charges, which are vectors of $O(6,n)$ and together  form
a doublet of $SL(2,Z)$. It is sometimes convenient
to write the charges as 
$V_{\alpha i} = (Q_i, P_i)$ where $\alpha =1,2$ is the $SL(2,Z)$
index. 
We can form the symmetric matrix $M_{\alpha \beta} =
V_{\alpha i } V_{\beta j } \eta^{ij}$ where 
$\eta^{ij}$ is the $O(6,n)$ metric. 
The black hole entropy is proportional to \cveticfifth \sentorus \duff 
\eqn\nfouren{
S
= \sqrt{ det(M) } = \sqrt{ Q^2 P^2 -(Q.P)^2 }
}
 which is clearly invariant. 

The condition for having a BPS solution is that $M_{\alpha \beta}$ is
 semi-definite positive which means that $ det M \ge 0$ and $M_{11}
\ge 0$, which  implies, in particular,  that $Q^2 \ge 0$ and $P^2 \ge 0$.

The condition for having a 1/2 BPS solution is that
$ \epsilon^{\alpha \beta} V_{\alpha i} V_{\beta j} = 0$ which 
means that $Q$ and $P$ are parallel vectors, so that by means
of an $SL(2,Z)$ transformation the configuration can be dualized into
one with only electric (or only magnetic) charges.
We can present this statement, in analogy to \projectadj  ,
by saying that the projection of the second derivatives of
the invariant $det M $ projected on the adjoint of $O(6,n)$ 
vanishes. As in $N=8$, there is generically a 
phase that cannot be removed. This is a phase between the central and
matter charges, reflecting the fact that  five parameters 
are necessary to obtain 
 the general solution \cveticfifth . Again this phase automatically
vanishes if we have a 1/2 BPS state.

{\bf Acknowledgements}

We thank M. Cvetic, F. Larsen and P. Pouliot  for discussions.
S. Ferrara would like to thank, for its kind hospitality, the 
 Physics Department of Rutgers
University where part of this work was done. 

This work was supported in part by EEC under TMR contract
ERBFMRX-CT96-00 (LNF Frascati, INFN, Italy)
 and DOE grants DE-FG03-91ER40662,
DE-FG02-96ER40559.

\listrefs

\bye